\documentclass[letterpaper, 10 pt, conference]{ieeeconf}  %

\IEEEoverridecommandlockouts                              %

\usepackage{cite}
\usepackage{amsmath,amssymb,amsfonts}
\usepackage{algorithmic}
\usepackage{graphicx}
\usepackage{textcomp}
\usepackage{xcolor}

\def\BibTeX{{\rm B\kern-.05em{\sc i\kern-.025em b}\kern-.08em
    T\kern-.1667em\lower.7ex\hbox{E}\kern-.125emX}}

\usepackage{tikz}
\usepackage{pgfplots}
\usepackage[per-mode=fraction, detect-weight=true]{siunitx}
\usepackage[hidelinks]{hyperref}
\usepackage{subcaption}
\usepackage{algorithm}

\usepgfplotslibrary{groupplots}
\usetikzlibrary{backgrounds}

\pgfplotsset{
	compat=1.18,
	standard-plot/.style={
		grid=major,
		grid style={dashdotted},
	},
	line-plot/.style={
		width=\columnwidth,
		height=0.5\columnwidth,
	},
	xy-plot/.style={
		width=\columnwidth,
		height=0.5\columnwidth,
		standard-plot,
	},
	group-plot/.style={
		width=\columnwidth,
		height=0.5\columnwidth,
	},
	group-plot-2/.style={
		group-plot,
		group style={
			group size=1 by 2,
			xlabels at=edge bottom,
			xticklabels at=edge bottom,
			vertical sep=0cm,
		},
	},
	group-plot-3/.style={
		group-plot,
		group style={
			group size=1 by 3,
			xlabels at=edge bottom,
			xticklabels at=edge bottom,
			vertical sep=0cm,
		},
	},
	group-plot-4/.style={
		group-plot,
		group style={group size=1 by 4,
			xlabels at=edge bottom,
			xticklabels at=edge bottom,
			vertical sep=0cm,
		},
	},
	first-group-plot/.style={
		standard-plot,
		enlarge x limits=0,
		xmin=0,
		y label style={yshift=-0.6em},
		legend columns=2,
		legend style={
				legend pos=north west,
			},
	},
	middle-group-plot/.style={
		standard-plot,
		enlarge x limits=0,
		xmin=0,
		y label style={yshift=-0.6em},
	},
	last-group-plot/.style={
		standard-plot,
		xlabel=$t$(\si{\second}),
		enlarge x limits=0,
		xmin=0,
		y label style={yshift=-0.6em},
	},
	generic-linestyle/.style={thick},
	reference/.style={generic-linestyle, TUMBlack},
	line-1/.style={generic-linestyle, TUMOrange},
	line-2/.style={generic-linestyle, TUMBlue},
	line-3/.style={generic-linestyle, TUMGreen},
	line-max/.style={generic-linestyle, TUMOrange, dashed},
	line-actual/.style={generic-linestyle, TUMBlue},
}

\usetikzlibrary{shapes}
\usetikzlibrary{positioning}
\usetikzlibrary{pgfplots.statistics}

\tikzset{
	flow-chart/.style={
			fc-block/.style={minimum height=1cm, minimum width=3.5cm, draw=TUMBlack, align=center},
			process/.style={fc-block, rectangle},
			process-predef/.style={fc-block, predproc},
			decision/.style={fc-block, diamond, aspect=2},
			flow-fusion/.style={decision, minimum width=1cm, aspect =1},
			input-output/.style={fc-block, trapezium, trapezium left angle = 65,trapezium right angle = 115,trapezium stretches},
			termination/.style={fc-block, rounded rectangle},
			flow/.style={-latex},
			flow-label/.style={fill=white, pos=0.5}
		},
	control-architecture/.style = {
			standard-block/.style={
					rectangle,
					rounded corners=.2cm,
					minimum height=1cm,
					minimum width=2cm,
					text width=2.5cm,
					inner xsep = 0cm,
					draw=TUMBlack,
					align=center,
				},
			software-block/.style={
					standard-block,
					draw=TUMBlue
				},
			software-block/.style={
					standard-block,
					white,
					draw=TUMBlue,
					fill=TUMBlue,
				},
			data-block/.style={
					standard-block,
				},
			standard-arrow/.style={-latex},
			standard-arrow-label/.style={fill=none, pos=0.5, align=left, text width =1.5cm},
			left-arrow-label/.style={fill=none, pos=0.5, align=right, text width =1.5cm}
		}
}

\definecolor{TUMBlue}{RGB}{0,101,189}%
\definecolor{TUMWhite}{RGB}{255,255,255}%
\definecolor{TUMBlack}{RGB}{0,0,0}%
\definecolor{TUMBlue1}{RGB}{0,51,89}%
\definecolor{TUMBlue2}{RGB}{0,82,147}%
\definecolor{TUMGray1}{RGB}{51,51,51}%
\definecolor{TUMGray2}{RGB}{127,127,127}%
\definecolor{TUMGray3}{RGB}{204,204,204}%
\definecolor{TUMBlue3}{RGB}{100,160,200}%
\definecolor{TUMBlue4}{RGB}{152,198,234}%
\definecolor{TUMIvory}{RGB}{218,215,203}%
\definecolor{TUMOrange}{RGB}{227,114,34}%
\definecolor{TUMGreen}{RGB}{162,173,0}%

\newlength{\nodewidth}
\newlength{\nodespacinghor}
\newlength{\nodespacingver}
\newlength{\arrowlength}
\pgfplotsset{
	compat=1.18,
	standard-plot/.style={
		grid=major,
		grid style={dashdotted},
	},
	line-plot/.style={
		width=\columnwidth,
		height=0.5\columnwidth,
	},
	xy-plot/.style={
		width=\columnwidth,
		height=0.5\columnwidth,
		standard-plot,
	},
	group-plot/.style={
		width=\columnwidth,
		height=0.5\columnwidth,
	},
	group-plot-2/.style={
		group-plot,
		group style={
			group size=1 by 2,
			xlabels at=edge bottom,
			xticklabels at=edge bottom,
			vertical sep=0cm,
		},
	},
	group-plot-3/.style={
		group-plot,
		group style={
			group size=1 by 3,
			xlabels at=edge bottom,
			xticklabels at=edge bottom,
			vertical sep=0cm,
		},
	},
	group-plot-4/.style={
		group-plot,
		group style={group size=1 by 4,
			xlabels at=edge bottom,
			xticklabels at=edge bottom,
			vertical sep=0cm,
		},
	},
	first-group-plot/.style={
		standard-plot,
		enlarge x limits=0,
		xmin=0,
		y label style={yshift=-0.6em},
		legend columns=2,
		legend style={
				legend pos=north west,
			},
	},
	middle-group-plot/.style={
		standard-plot,
		enlarge x limits=0,
		xmin=0,
		y label style={yshift=-0.6em},
	},
	last-group-plot/.style={
		standard-plot,
		xlabel=$t$(\si{\second}),
		enlarge x limits=0,
		xmin=0,
		y label style={yshift=-0.6em},
	},
	generic-linestyle/.style={thick},
	reference/.style={generic-linestyle, TUMBlack},
	line-1/.style={generic-linestyle, TUMOrange},
	line-2/.style={generic-linestyle, TUMBlue},
	line-3/.style={generic-linestyle, TUMGreen},
	line-max/.style={generic-linestyle, TUMOrange, dashed},
	line-actual/.style={generic-linestyle, TUMBlue},
}

\title{\LARGE \bf
Longitudinal Control for Autonomous Racing with Combustion Engine Vehicles
\thanks{$^{1}$Technical University of Munich, Germany; School of Engineering \& Design, Department of Engineering Physics and Computation, Institute of Automatic Control}
\thanks{$^{2}$Technical University of Munich, Germany; School of Engineering \& Design, Department of Mobility Systems Engineering, Institute of Automotive Technology\newline{}
	Corresponding author: \href{mailto:phillip.pitschi@tum.de}{phillip.pitschi@tum.de}}
}

\author{Phillip Pitschi$^{1}$, Simon Sagmeister$^{2}$, Sven Goblirsch$^{2}$, Markus Lienkamp$^{2}$, Boris Lohmann$^{1}$}%

\newcommand\copyrighttext{%
	\footnotesize \textcopyright 2025 IEEE.  Personal use of this material is permitted.  Permission from IEEE must be obtained for all other uses, in any current or future media, including reprinting/republishing this material for advertising or promotional purposes, creating new collective works, for resale or redistribution to servers or lists, or reuse of any copyrighted component of this work in other works.
}

\newcommand\copyrightnotice{%
	\begin{tikzpicture}[remember picture,overlay]
	\node[anchor=south,yshift=10pt, xshift=0pt] at (current page.south) {\fbox{\parbox{\dimexpr\textwidth-\fboxsep-\fboxrule\relax}{\copyrighttext}}};
	\end{tikzpicture}%
    \vspace{-0.35cm}
}

\begin{document}

\maketitle
\thispagestyle{empty}
\pagestyle{empty}
\copyrightnotice{}

\setcounter{footnote}{1}
\begin{abstract}
    Usually, a controller for path- or trajectory tracking is employed in autonomous driving. Typically, these controllers generate high-level commands like longitudinal acceleration or force. However, vehicles with combustion engines expect different actuation inputs. This paper proposes a longitudinal control concept that translates high-level trajectory-tracking commands to the required low-level vehicle commands such as throttle, brake pressure and a desired gear. We chose a modular structure to easily integrate different trajectory-tracking control algorithms and vehicles. The proposed control concept enables a close tracking of the high-level control command. An anti-lock braking system, traction control, and brake warmup control also ensure a safe operation during real-world tests. We provide experimental validation of our concept using real world data with longitudinal accelerations reaching up to \SI{25}{\meter\per\second\squared}. The experiments were conducted using the EAV24 racecar during the first event of the Abu Dhabi Autonomous Racing League on the Yas Marina Formula 1 Circuit.
\end{abstract}

\section{Introduction}

Controlling an autonomous vehicle is particularly challenging, especially in racing scenarios, where the vehicle has to adhere to a target trajectory or path while maintaining stability at its dynamic limits. Numerous trajectory- and path-tracking control approaches have been developed to address this issue. However, most of these methods only provide a high-level control output, including a steering angle and either longitudinal acceleration or force. These high-level longitudinal commands can not be used directly to control a combustion engine vehicle. Therefore, applying a tracking controller to such vehicles requires a \textit{longitudinal control system} that translates these to compatible actuation commands. For our testing vehicle, these are throttle, brake pressure, and a gear request.

Additionally, a performant longitudinal control system is required to \textit{evaluate} and \textit{compare} different trajectory- and path-tracking controllers on real-world vehicles or in high-fidelity simulations (including drivetrain dynamics). The overall control performance is highly affected by the tracking accuracy of the high-level commands. A poor tracking of longitudinal acceleration or force would complicate the comparisons between trajectory- and path-tracking controllers.

Moreover, ensuring the \textit{safety} and \textit{stability} of a vehicle during real-world tests is essential. Particularly in racing, where vehicles operate at their dynamic limits, they can quickly become unstable. If the tires on one axle lock or spin, the vehicle can swerve and potentially crash. To mitigate this risk, the incorporation of additional safety features is necessary.

\subsection{Research Vehicle}

\begin{figure}[!t]
	\centering
	\includegraphics[width=\columnwidth]{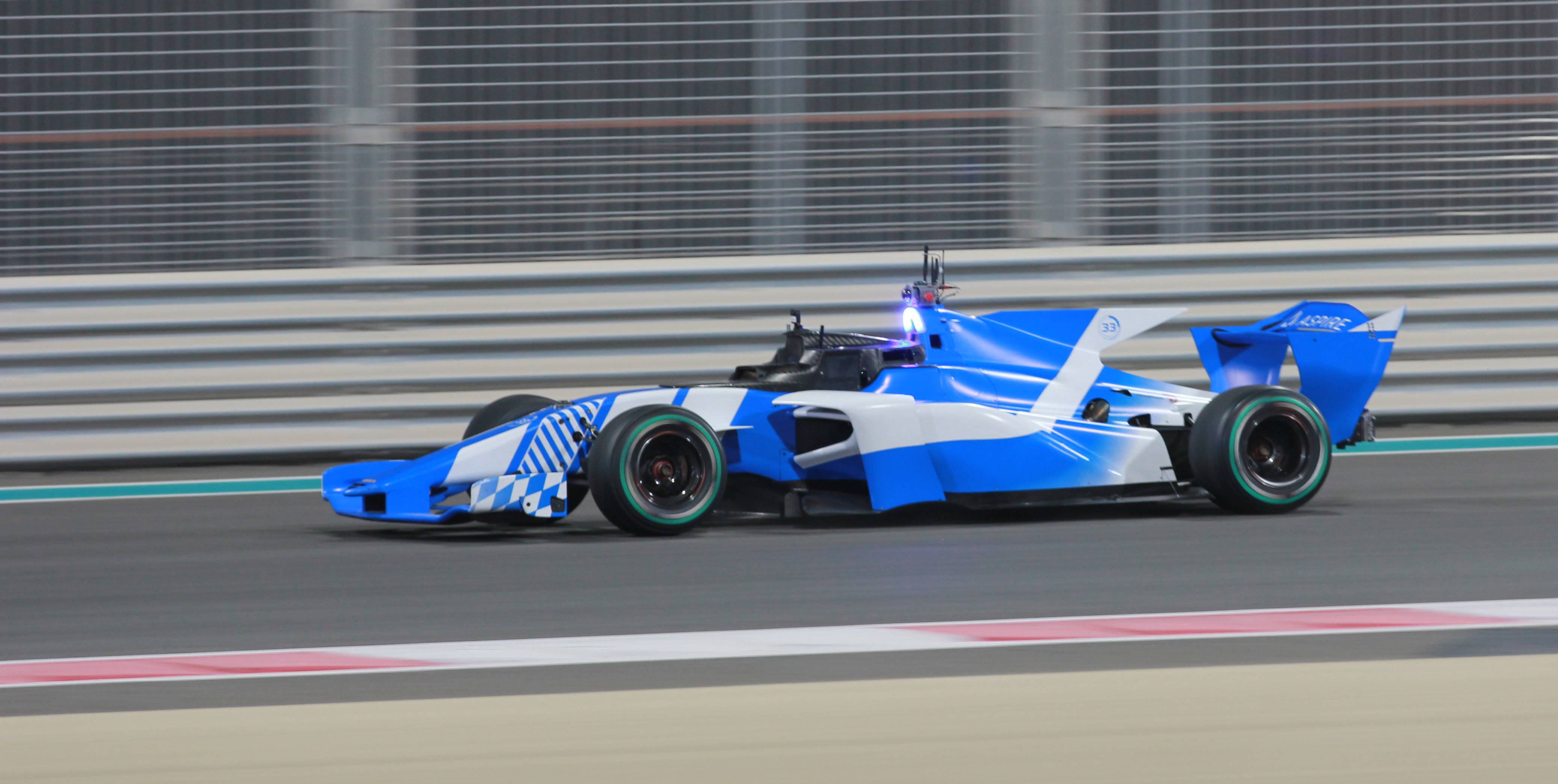}
	\caption{Dallara EAV24 racecar of the TUM Autonomous Motorsport team at Yas Marina Circuit in Abu Dhabi}
	\label{fig:Hailey}
\end{figure}

We deployed the longitudinal control system presented in the remainder of this work during our participation at the Abu Dhabi Autonomous Racing League.
Fig. \ref{fig:Hailey} shows the Dallara EAV24 racecar used in this competition. It is based on the SF23 Super Formula Chassis made by Dallara.

The car has a turbocharged combustion engine with a sequential gearbox. However, the driver's seat was removed and replaced with an autonomous driving stack to actuate the vehicle.
Sending target values for throttle and gear allows the control of the drivetrain from an autonomous software stack.
Similarly, a target steering angle and a brake pressure for each wheel individually can be commanded.
Like the original SF23, the autonomous race car has state-of-the-art carbon-ceramic brake discs.

\subsection{Challenges}

However, it is these \textit{carbon-ceramic brakes} that pose one of the particular challenges. The friction coefficients of carbon brake disks vary with temperature. For the braking system to work consistently, the brake disks must be in an optimal temperature range~\cite{Yevtushenko2020, Goeltz2020}.

Additionally, numerous factors influence the \textit{friction at the tire-road contact}, including tire and road temperature, track conditions, and tire wear. Modeling the interaction between tire and road is highly complex, and an open research topic~\cite{Woodward2012, Waddell2019}. For the safe operation of a real-world vehicle, locking or spinning of tires has to be prevented under all driving conditions.

The previously introduced racecar features a \textit{turbocharged combustion engine}. This highly varies in the torque output depending on oxygen levels, air temperature, and boost. Consequently, a static mapping from engine speed and desired torque to throttle constitutes an only suboptimal solution~\cite{Lefebvre2005}.

Moreover, combustion engines exert a \textit{drag torque} on a vehicle for low throttle values. A controller has to account for this additional torque in braking situations, necessitating adjustments of the brake pressure commands.

A comprehensive concept that addresses these challenges has not yet been introduced in the context of vehicle dynamics control. Hence, this paper comprises the following contributions:
\begin{itemize}
	\item We propose a novel concept for a performant and safe longitudinal control system of a vehicle up to its handling limits. This approach enables the safe testing of various trajectory- and path-tracking control algorithms on real-world vehicles. The developed algorithm generates the same inputs that a human driver would employ, making the concept applicable to all vehicles operable by a human.
	\item We present an experimental validation of our approach executed on a real-world race car. The results show that the developed controller can safely handle the vehicle in its complete operation area.
	\item We provide the developed controller as open-source software at \url{https://github.com/TUMFTM/TAM__long_acc_control} to facilitate real-world testing and comparison of trajectory- and path-tracking control methods. In order to simplify integration, the controller is compatible with the intefaces defined by Autoware.
\end{itemize}

\section{Related Work}

Multiple publications address the issue of longitudinal control of vehicles. This includes both pure longitudinal controllers~\cite{Pedone2020, Ni2019, Laurense2018}, and combined control of lateral and longitudinal motions~\cite{Vazquez2020, Wohner2021, Kalaria2024, Numerow2024, Costa2023}. Paden et al.~\cite{Paden2016} and Betz et al.~\cite{Betz2022} provide a comprehensive overview of diverse control concepts applicable to urban and racing scenarios, respectively. However, these methods provide only high-level control commands.

The different trajectory- and path-tracking control concepts include classic approaches such as proportional control~\cite{Ni2019}, slip angle-based control~\cite{Laurense2018}, and model predictive control~\cite{Vazquez2020, Wohner2021, Kalaria2024, Numerow2024, Costa2023}. These control approaches produce a driving torque~\cite{Pedone2020, Wohner2021, Costa2023}, longitudinal force~\cite{Ni2019, Laurense2018, Vazquez2020, Numerow2024} or longitudinal acceleration~\cite{Kalaria2024} as control commands. These must be translated to brake pressure, throttle, and gear. Furthermore, none of the approaches account for the varying friction coefficients of tires or brake discs or drag torque introduced by a combustion engine.

The Autoware Foundation offers a vehicle interface that maps a high-level acceleration command to low-level commands needed to control a real-world car. This interface includes a state machine that checks command ranges and oscillations but does not specify how to match the high-level command to a brake pressure and throttle command. Additionally, it assumes the presence of an automatic gearbox and does not support manual gear changes~\cite{kato2018, theautowarefoundation}.

A few publications exist regarding control concepts for autonomous racing with combustion engine vehicles. The trajectory-tracking control methods used in this context include model predictive control~\cite{Raji2022, Raji2023, Xue2023, Pierini2024, Wischnewski2023}, linear quadratic regulator~\cite{Chung2024} and pure pursuit control~\cite{Sukhil2021}. These approaches typically output either a longitudinal acceleration~\cite{Raji2023, Chung2024, Xue2023, Wischnewski2023} or a longitudinal force~\cite{Raji2022, Pierini2024}. Pierini et al.~\cite{Pierini2024} utilize an inverted engine map lookup table and brake characteristics for a feedforward calculation of the low-level control commands. Chung et al.~\cite{Chung2024} combine such a feedforward structure with a feedback PID controller, while Raji et al.~\cite{Raji2023} additionally consider driving resistances impacting the vehicle.

However, none of these concepts explicitly account for the engine's drag torque or varying brake disc friction. Raji et al.~\cite{Raji2023} limit the longitudinal force when calculating brake pressure and throttle using a pacejka tire model. Nevertheless, this model needs an accurate estimate of the tire friction coefficient, which is usually chosen constant because of the lack of more suitable information. If an unexpected change in the friction occurs, this controller cannot react accordingly.

Consequently, none of the existent control concepts provide a performant and safe longitudinal control method capable of addressing all challenges arising when a real-world racecar operates at its dynamic limits.

\section{Methodology}

\begin{figure}
	\centering
	\includegraphics{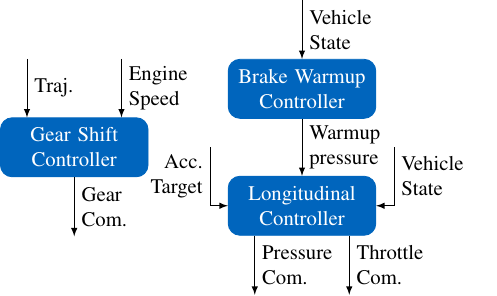}
	\caption{Structure of the longitudinal control system}
	\label{fig:Structure}
\end{figure}

The structure of the proposed longitudinal control system is depicted in Fig.~\ref{fig:Structure}. It consists of a gear shift controller, a brake warmup controller, and a longitudinal controller. The gear shift controller produces a gear command based on the current trajectory and engine speed. The brake warmup controller is an upstream module for the longitudinal controller. It provides additional brake pressure based on the vehicle's current state in order to accelerate brake disc warmup. The longitudinal controller converts a high-level acceleration target to a brake pressure and throttle command using the current vehicle state.

We design each of these modules as separate ROS2 nodes. This way, the three modules can be executed at different frequencies to save CPU resources. The gear shift and brake warmup controllers run at a lower frequency than the longitudinal controller, which has to rapidly adjust the throttle and brake pressure commands to ensure vehicle stability and close tracking of the acceleration target. Furthermore, we can easily disable the gear shift module for vehicles with an automatic gearbox. Similarly, we achieve flexibility in using the brake warmup controller for vehicles with different braking systems.

\subsection{Predictive Gear Shift Controller}

The torque curve of combustion engines usually exhibits a plateau of maximum available torque in a specific engine speed range. Racecar drivers, therefore, need to ensure the use of the correct gear to maximize the vehicle's acceleration. Segers~\cite{segers2014} analyzes the impact of upshifting and downshifting strategies for different drivers. The results show a high impact on the final lap time, thus emphasizing the importance of a well-designed gear shift strategy.

Besides selecting the correct gear, shift timing is crucial. A gear shift during cornering can decrease tire adhesion and induce snap oversteer~\cite{segers2014}. We cut the gear shift request to prevent this if the lateral acceleration exceeds a specific limit. This acceleration limit is set based on a velocity and vertical acceleration dependent gg-diagram - called gggv-diagram~\cite{Rowold2023}. Additionally, we introduce a predictive gear shift strategy to anticipate the need for an early gear shift. This strategy allows maximum torque output at the turn exit and avoids hitting the rev limiter when a shift is not permissible.

We utilize the planned trajectory to predict the target velocity and, thus, the corresponding engine speed with the current gear. We evaluate whether a downshift will be necessary based on this engine speed. We use a lookahead horizon starting after a specific time delay $t_{\text{lookahead}}$ to account for oscillations induced by a gear shift. If the planned lateral acceleration exceeds the previously introduced limit at every point between this time delay and the point at which the lower gear is required $t_{\text{shift}}$, an early shift is performed.

\begin{figure}
	\centering
	\includegraphics{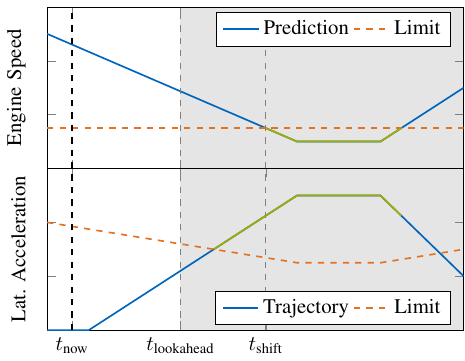}
	\caption{Predictive gear shift controller functionality. The gray shaded area indicates the lookahead horizon. The green parts mark the region where a shift is required but not executed because of high lateral accelerations.}
	\label{fig:GSCFunctionality}
\end{figure}

Fig.~\ref{fig:GSCFunctionality} illustrates the functionality of the predictive gear shift controller. A shift will be required when the engine speed drops below the downshift limit. For the current time, the lateral acceleration is still below the limit at the start of the lookahead horizon. We perform an early shift once the horizon reaches the point where the acceleration exceeds the limit. Note that the lateral acceleration limit decreases with lower vehicle velocities. To prevent excessive drag torque and over-revving the engine, we do not perform a shift in case a certain engine speed is exceeded with the lower gear.

\subsection{Brake Warmup Controller}

As mentioned, carbon-ceramic brakes must be in a specific temperature window to work properly.
Therefore, it is essential to heat the brakes as fast as possible to enable time-efficient testing.
To deal with this issue, we created a brake warmup controller that applies a small amount of brake pressure during normal driving.
In order to not negatively influence the vehicle in corners, we deactivate the brake warmup controller based on the lateral acceleration target.
Furthermore, the brake warmup controller will be disabled after a preprogrammed number of laps or after the brakes have reached a specific temperature.
As shown in Fig.~\ref{fig:Structure}, the brake warmup controller provides the warmup pressure as an - for the longitudinal controller optional - input to the longitudinal controller.
In the longitudinal controller, the maximum of the original brake pressure target and the warmup pressure is used as the final brake pressure command.
This way, we ensure that the brake warmup controller does not interfere with the vehicle's braking behavior.
The brake warmup controller enabled us to heat the brakes within one lap on the Yas Marina Circuit up to their operating temperature of around \SI{450}{\degreeCelsius}.

\subsection{Longitudinal Controller}

\begin{figure}
	\centering
	\includegraphics{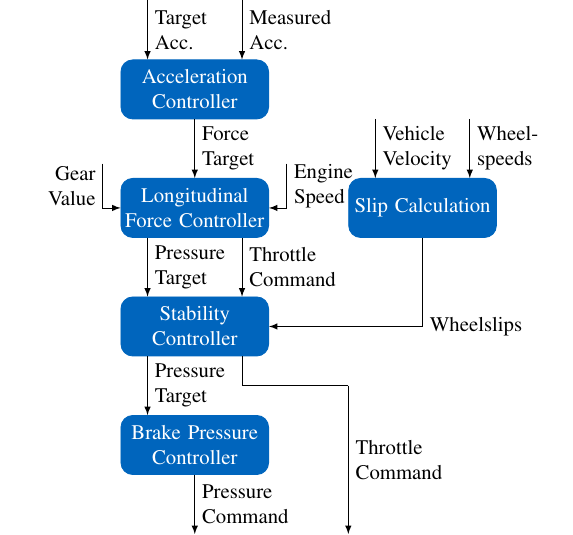}
	\caption{Internal structure of the longitudinal controller}
	\label{fig:LongControlStructure}
\end{figure}

The longitudinal controller converts an acceleration or force into a throttle and brake command, ensuring accurate acceleration tracking and vehicle stability. The internal structure of the longitudinal controller and its subcomponents is depicted in Fig.~\ref{fig:LongControlStructure}.

The acceleration controller compares the measured acceleration and an acceleration target produced by an arbitrary trajectory-tracking controller and generates a longitudinal force target. The longitudinal force controller converts this force target to brake pressure and throttle targets using the currently engaged gear and engine speed. The stability controller adjusts these brake pressure and throttle targets to ensure vehicle stability. To achieve this, the stability controller receives the current wheel slips from the slip calculation component. This computes the current wheel slips based on the vehicle's dynamic state and the individual wheel speeds. The adjusted throttle value is used directly as a command. At the same time, the brake pressure controller processes the brake pressure target along with the measured brake pressure to generate the final brake command.

\subsubsection{Acceleration Controller}

The acceleration controller consists of a feedforward control and a feedback PID control part. The feedforward part generates a target force,

\begin{equation}
	F_{x,\mathrm{target}} = \left( m + \frac{J_{\mathrm{drivetrain}}}{r_{\mathrm{wheel},r}^2} \right) \, a_{x,\mathrm{target}} + F_{\mathrm{aero}} + F_{\mathrm{roll}},
\end{equation}

that combines the trajectory-tracking controller's acceleration target $a_{x,\mathrm{target}}$ and components representing driving resistances. We scale the acceleration target with the vehicle mass $m$ and a term that includes the drivetrain's moment of inertia $J_{\mathrm{drivetrain}}$ and the rear wheel radius $r_{\mathrm{wheel},r}$. The driving resistances are modeled as aerodynamic drag $F_{\mathrm{aero}}$ and rolling resistance $F_{\mathrm{roll}}$.

The feedback part regulates the error between the target and measured acceleration value in the longitudinal direction.

\subsubsection{Longitudinal Force Controller}

The longitudinal force controller calculates a motor torque and brake pressure target from the longitudinal force $F_{x,\mathrm{target}}$, converts the motor torque to a throttle target, and regulates the torque to match the target and measured acceleration.

To calculate the motor torque target $T_{\mathrm{motor},\mathrm{target}}$ from the longitudinal force target, the controller uses the ratio of the gearbox $\tau_{\mathrm{gearbox}}$ at the currently engaged gear, the ratio of the final drive $\tau_{\mathrm{final\,drive}}$, the drivetrain efficiency $\eta_{\mathrm{drivetrain}}$ and wheel radius at the rear axle:
\begin{equation}
	T_{\mathrm{motor},\mathrm{target}} = \frac{r_{\mathrm{wheel},r}}{\tau_{\mathrm{gearbox}} \, \tau_{\mathrm{final\,drive}} \, \eta_{\mathrm{drivetrain}}} \, F_{x,\mathrm{target}}.
\end{equation}

We add additional terms to avoid stalling or over-revving the engine. With the requested motor torque and the current engine speed, a throttle target is queried from a lookup table that maps the static motor characteristics. An additional PID controller with an anti-wind-up mechanism regulates the throttle command to match the measured longitudinal acceleration to the target.

We calculate the target brake pressure for the front or rear axle,

\begin{equation}
	p_{b,\mathrm{target},i} = \zeta_i \, \frac{r_{\mathrm{wheel},i}}{2 \, \pi \,  \left( \frac{d_{\mathrm{bore}}} {2} \right)^2 \, \mu_{\mathrm{brake}} \, r_{\mathrm{lever}} } \, F_{x,\mathrm{brake}},
\end{equation}

based on the braking system's geometry and characteristics, including the wheel radius $r_{\mathrm{wheel},i}$, bore diameter $d_{\mathrm{bore}}$, the friction coefficient $\mu_{\mathrm{brake}}$ of the brake pads and the mean brake lever radius $r_{\mathrm{lever}}$. The brake force $F_{x,\mathrm{brake}}$ is calculated by subtracting the braking force resulting from the engine drag torque from the longitudinal force target. The brake bias $\zeta_i$ distributes the brake force between the front and rear axle. As the drag torque, in our case, only acts on the rear axle, we adjust the brake bias to allocate the braking force according to a fixed ratio.

\subsubsection{Slip Calculation}
We calculate exact slip values for each wheel, as we need these as input to the downstream stability controller (Fig.~\ref{fig:LongControlStructure}).
To calculate the slip values, we use the established correlations for a two-track model~\cite{allen1992} and the slip definition according to Pacejka~\cite{pacejka2012}. In addition, there are two specializations for our application:

Due to the high speeds reached by the racing car, we have to consider the speed-dependency of the effective tire rolling radius during slip calculation. The correlation between speed and tire rolling radius is stored in a calibration curve. This enables simple parameterization from real-world data as well as data sheets.

Since the race car is rear-wheel driven only, we assume the front wheels do not slip during acceleration. From this assumption, we can calculate a velocity at the car's center of gravity by inverting the equations used during conventional slip calculation.
This method results in a accurate estimation of the vehicle's velocity and increases robustness against disturbances in state estimation. However, because the zero slip assumption is not valid under braking, we must smoothly transition from the front wheel-based velocity to the state estimation velocity.
We calculate an interpolation factor between the front wheel velocity ($k_\mathrm{\kappa} = 0$) and the state estimation velocity ($k_\mathrm{\kappa} = 1$):
\begin{equation}
	\begin{aligned}
		k_\mathrm{\kappa} = \max\left(\frac{p_{\mathrm{b,\,f}}}{\SI{5e5}{\pascal}}, -\left(a_{x,\,\mathrm{filtered}} + \SI{5}{\meter\per\second\squared}\right) \right).
	\end{aligned}
\end{equation}
The interpolation factor $k_\mathrm{\kappa}$ is limited to an interval of $[0,1]$ and depends on the brake pressure of the front wheels $p_{\mathrm{b},\mathrm{f}}$ as well as the filtered longitudinal acceleration $a_{x,\,\mathrm{filtered}}$. By using both acceleration and brake pressure as indicators for deceleration, we increase robustness against sensor failures. To ensure smooth transitions between velocity estimates even in highly dynamic situations, we use a low-pass filtered representation of $k_\mathrm{\kappa}$.

\subsubsection{Stability Controller}

In autonomous racing, the accelerations approach the limits of the vehicle's capabilities. If the applied torque at the wheels surpasses the tire's capabilities due to a wrong estimate of tire friction, the wheels can lock or spin. To mitigate this risk, we employ an anti-lock braking system (ABS) and traction control (TC) for decelerating and accelerating. (In contrast to conventional road vehicles, our research vehicle does not have a brake actor designed to apply an ABS or TC. Therefore, standard algorithms from production vehicles are not applicable.)

\begin{figure}
	\centering
	\includegraphics{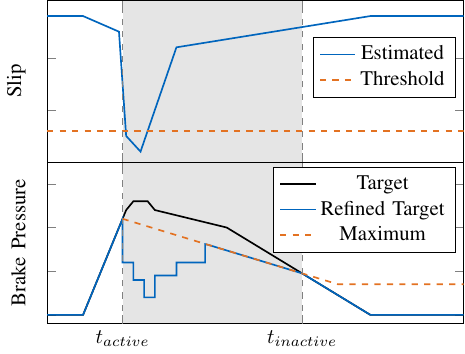}
	\caption{ABS functionality. The gray shaded area indicates the active ABS intervention.}
	\label{fig:ABSFunctionality}
\end{figure}

We use a heuristic approach for both the ABS and TC, wherein the systems adjust the brake pressure target for each axle individually. Fig.~\ref{fig:ABSFunctionality} illustrates the functionality of the ABS, showing qualitative courses of the estimated wheel slip and the brake pressure target for an exemplary ABS intervention.

As long as the system is inactive, it forwards the target value it receives from the upstream components. The ABS starts its intervention if any of the wheels slips exceed a predefined threshold. The system latches its brake pressure target at activation as the maximum permissible brake pressure. During the braking maneuver, we gradually decrease this maximum value. This reduction accounts for the diminishing vertical tire force due to aerodynamic effects for decreasing velocities during a braking maneuver. When the brake pressure target received from the upstream modules falls below the current maximum permissible brake pressure, we terminate the ABS intervention, returning to regular operation.

During an ABS intervention, we reduce the brake pressure target relative to the latched value. Upstream modules include integral parts that try to compensate for this reduction by increasing the brake pressure target. Therefore, we calculate the refined target as a ratio of the current maximum value that is independent of the received target after ABS activation.

The ratio decreases if the slip exceeds the defined threshold and increases if the slip is below the threshold. We decrease the ratio from \SI{100}{\percent} to a predefined value during the initial step. We perform relative percentage steps for all subsequent steps depending on the distance between the slip value and threshold.

For the ABS, we introduce an additional safety feature in case the system reduces the brake pressure target to a low value for a prolonged period, hindering the vehicle from decelerating to the desired velocity. The increased velocity can lead to infeasible lateral accelerations in an upcoming turn. Therefore, we pause the ABS intervention if the resulting longitudinal force falls below a certain threshold for a predefined time. This feature is not added to the TC, as it is not safety-critical if the vehicle drives at a slower velocity than desired.

The TC system works analogously to the ABS. However, we do not have direct access to the engine's control unit and can only send a throttle command to it. This command influences the engine's throttle valve but does not directly affect the ignition system. Consequently, in comparison to the response time of the brake actuators, the motor torque changes only slowly for an adjusted throttle target. Therefore, we transform the required drive torque reduction into an increase of the brake pressure target. While the throttle target remains at the desired value, the engine torque is superimposed with the brake torque at the wheels, allowing for rapid adjustments of the driving torque. In contrast to the ABS, we keep the maximum value constant in the TC as the velocity increases during the intervention.

Additionally, if the slip exceeds an upper threshold, slip stabilization is unlikely solely with an increased brake pressure target. In this case, we reduce the throttle target to zero.

\subsubsection{Brake Pressure Controller}

The measured brake pressure must closely follow the target for a good acceleration following behavior and for the ABS and TC concepts to work as intended. To improve the setpoint tracking and dynamic response of the braking actuator system, we introduce a cascaded Proportional-Integral (PI) controller. This controller is designed to track the desired brake pressure by comparing it to the measured brake pressure. The PI system is further enhanced with a smith prediction to handle feedback time delays, gain scheduling to adjust the control dynamics based on the brake pressure gradient, and an anti-wind-up mechanism.

\section{Results}

The results shown in this section originate from the first event of the Abu Dhabi Autonomous Racing League on the Yas Marina Formula 1 Circuit. We evaluate the longitudinal control system in combination with a slightly modified version of the trajectory-tracking controller developed by Wischnewski et al.~\cite{Wischnewski2023}.

\subsection{Testing Strategy}
After implementing the features described in this work, testing them to evaluate their performance is essential.
However, directly acting on the vehicle's actuators can be very dangerous. Therefore, we first test every new development extensively in our high-fidelity simulation environment~\cite{sagmeister2024}. However, even after successful evaluation in simulation, testing, especially the stability controller, poses the risk of damaging the test vehicle.

Therefore, we first started the real-world testing of the ABS at low speeds and with a brake bias that would safely lock the front wheels first.
This way, the car would stay stable even if the developed stability controller failed. After initial verification at low speeds, we increased the speed step by step, moving to a more rearward brake bias.
To evaluate the developed TC system, we first started with straight-line acceleration tests. Afterward, we tested combined situations by accelerating hard out of turns with big runoff areas. \\
These strategies enabled us to safely explore the limits of the developed system without risking the vehicle.

\subsection{Tracking of Longitudinal Acceleration}

\begin{figure}
    \centering
    \includegraphics{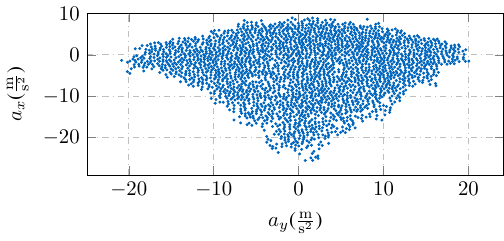}
    \caption{gg-Diagram for a testing session at the Yas Marina Circuit}
    \label{fig:GGDiagram}
\end{figure}

We first investigate the operation area of the longitudinal control system. For this, we look at the gg-diagram of a testing session at the Yas Marina Circuit, depicted in Fig.~\ref{fig:GGDiagram}. The lateral acceleration $a_y$ spans symmetrically to maximum absolute values of \SI{20}{\metre\per\square\second}. The longitudinal acceleration $a_x$ ranges from \SI{-25}{\metre\per\square\second} to \SI{8}{\metre\per\square\second}. The lower maximum absolute value of the accelerations in a positive longitudinal direction stems from the engine's limited power. In contrast, the negative accelerations are limited by the tire-road friction. The longitudinal control system can track the acceleration target for this operation area.

\begin{figure}
    \centering
    \includegraphics{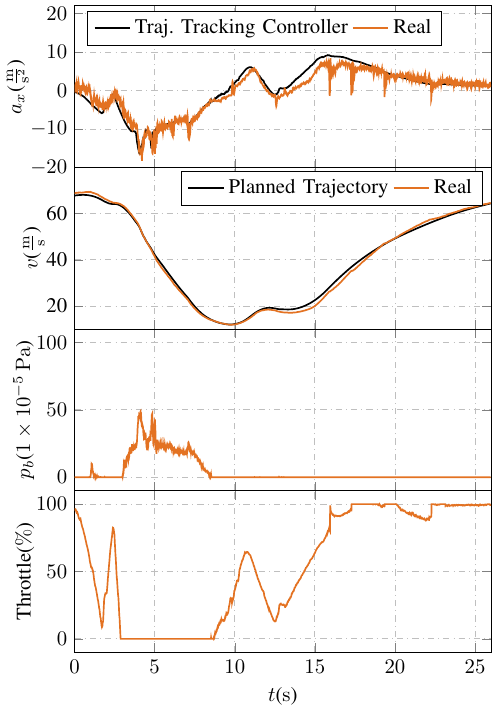}
    \caption{Acceleration and speed tracking and corresponding brake and throttle commands of the longitudinal control system. The longitudinal acceleration target corresponds to the trajectory-tracking controller command, whereas the velocity target stems from the trajectory planned by an upstream planning module.}
    \label{fig:AccControlPerformance}
\end{figure}

We now evaluate the acceleration tracking for a specific scenario. In this scenario, the vehicle drives at a high velocity towards a narrow chicane. The scenario consists of a braking phase before the entry to the first turn, a phase with only small accelerations between the two turns, and a long stretched acceleration phase after the second turn. We assess the longitudinal controller's isolated tracking of $a_y$ and the overall tracking of $v$ in combination with the trajectory-tracking controller with the courses displayed in Fig.~\ref{fig:AccControlPerformance}. Additionally, the commanded brake pressure $p_b$ and throttle commands are depicted.

During braking, the controller tracks the acceleration and velocity targets well. A slight velocity reduction is performed at the beginning of the braking maneuver. After that, the vehicle stays at the same velocity for a short period before the vehicle starts braking again. The initial velocity reduction is executed with only a low brake pressure. The central portion of the braking force stems from aerodynamic drag and the engine's drag torque. This shows the importance of integrating these effects into the longitudinal control system to command an appropriate brake pressure.

After the first turn at $t=\SI{11.8}{\second}$, the velocity target stays approximately constant; the actual velocity, however, decreases slightly. The trajectory-tracking controller reacts with a sudden increase of the acceleration target at $t=\SI{12.7}{\second}$. The longitudinal control system is not able to track this increase precisely. Instead, the acceleration only builds up slowly until $t=\SI{14.7}{\second}$ due to a delayed torque of the turbocharged engine. This introduces a small error in the velocity tracking remaining until $t=\SI{19}{\second}$. The throttle value is not at its limit for this situation, which suggests that a more aggressive reaction could lead to a smaller error. However, a corresponding tuning of the feedback controllers would also introduce an overshoot and oscillations in the acceleration tracking.

\subsection{Predictive Gear Shift Strategy}

\begin{figure}
    \centering
    \includegraphics{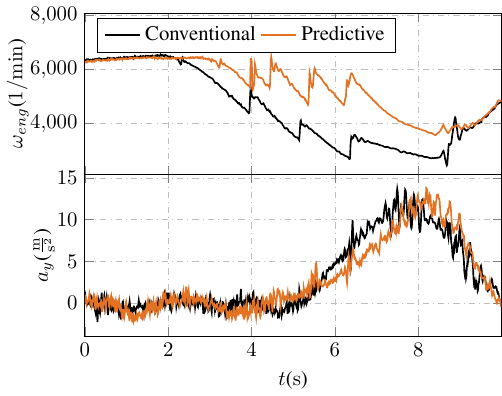}
    \caption{Comparison of engine speed and lateral acceleration for conventional and predictive gear shift strategies}
    \label{fig:GearShift}
\end{figure}

We evaluate the presented gear shift strategy in a low-speed corner after a long straight. Consequently, a high deceleration and, thus, multiple downshifts are required. We compare the gear shift strategy to a conventional strategy, where the gear is shifted down as soon as the engine speed drops below a certain threshold. Fig.~\ref{fig:GearShift} shows the vehicle's engine speed $\omega_{\mathrm{eng}}$ and lateral acceleration $a_y$ for both strategies. The spikes in the engine speed correspond to the gear shifts. As can be seen, the conventional strategy downshifts later than the predictive strategy. This leads to a shift shortly after lateral acceleration reaches its maximum value. As previously described, this can lead to a loss of traction. On the other hand, the predictive strategy enables the same engine speed at the corner exit and, thus, the optimal engine torque. However, the shifts are performed before the lateral acceleration reaches a critical limit.

\subsection{Braking Maneuver with ABS}

\begin{figure}
    \centering
    \includegraphics{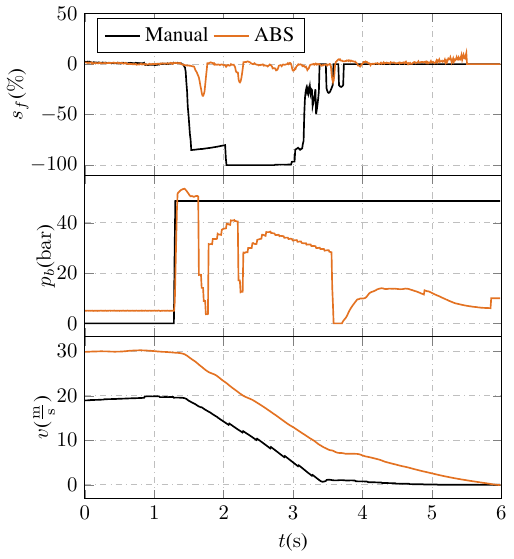}
    \caption{Comparison of front tire slip, brake pressure and velocity for a manual brake maneuver with joystick and an ABS controlled brake maneuver}
    \label{fig:ABSComparison}
\end{figure}

We compare two braking scenarios at low speeds. The first scenario is a manual brake maneuver using the joystick override, in which a constant brake pressure is applied abruptly. The second scenario is a braking maneuver with the proposed longitudinal control system, including the ABS. The wheel slip $s_f$ and brake pressure command $p_b$ for the front right wheel and the vehicle's velocity $v$ for both scenarios are depicted in Fig.~\ref{fig:ABSComparison}.

In the manual brake maneuver, the wheel slip increases to values close to \SI{100}{\percent} shortly after the brake pressure is applied. The wheel locks for approximately \SI{1.5}{\second} until the vehicle is already close to a standstill.

In the braking maneuver with the longitudinal control concept, the ABS can keep the wheel slip below a maximum absolute value of \SI{30}{\percent}. At $t=\SI{1.6}{\second}$ and $t=\SI{2.2}{\second}$ the slip starts to increase. The ABS reacts by reducing the brake pressure and can prevent the wheel from locking. The brake pressure increases to the maximum value after the slip value falls below the threshold again. After the second increase, the brake pressure stays at the maximum value until the target falls below the maximum at $t=\SI{3.6}{\second}$. The standstill is delayed to $t=\SI{5.9}{\second}$ due to an erroneous command of the upstream trajectory-tracking controller. The constant brake pressure at the beginning of the ABS brake maneuver corresponds to the applied warmup brake pressure.

The comparison shows that the ABS controller can prevent the wheel from locking, ensuring the vehicle's steerability during the complete braking maneuver.

\section{Conclusion}
\label{sec:conclusion}

We presented a modular longitudinal control system for autonomous vehicles with combustion engines. The approach translates high-level commands of a trajectory- or path-tracking controller into brake pressure, throttle, and gear commands. We showed that the controller can track longitudinal accelerations in highly dynamic scenarios. We demonstrated close tracking of the acceleration target for a deceleration situation. This was achieved by considering the engine's drag torque and driving resistance in the controller. However, in long-stretched acceleration situations, the controller struggles to regulate acceleration errors precisely. The introduced feedback controllers cannot fully compensate for the turbocharged engine's high momentum. In future work, we will introduce an additional data-based feedforward controller to represent these dynamic motor effects.

Additionally, we introduced a brake temperature controller, predictive gear shift algorithm, ABS, and TC to maximize safety in all situations during real-world experiments. We demonstrated that our predictive gear shift concept can avoid gear shifts in situations with high lateral accelerations while choosing the correct gear for maximum acceleration at the turn exit. Furthermore, we showed that the ABS can prevent the wheels from locking and ensures the vehicle's steerability during braking maneuvers. This guarantees the safety of the vehicle even for varying tire-road friction.

In future work, we want to extend the longitudinal control system with an electronic stability program. This additional module enables us to ensure the safety of the vehicle in oversteer situations when it starts to spin.

\section*{Acknowledgment}
Author contributions: Phillip Pitschi, as the first author, designed together with Simon Sagmeister the structure of the article and contributed essentially to the implementation as well as the contents of this paper.
Sven Goblirsch did the main development and implementation for the predictive gear shift strategy.
Boris Lohmann and Markus Lienkamp made an essential contribution to the concept of the research project.
We also thank Clemens Herrmann for the design, implementation, and simulative testing of the first version of the antilock-braking system.
Additionally, Frederik Werner designed the initial version of the brake pressure controller and contributed to the development and refinement throughout the overall control architecture during the last years of development. Tobias Betz contributed essentially to the implementation of the brake warmup controller.

\bibliographystyle{IEEEtran}
\bibliography{IEEEabrv,bibliography/saegi.bib, bibliography/phillip.bib, bibliography/sven.bib}

\end{document}